\documentclass[useAMS,usenatbib]{mn2e}

\usepackage{graphicx}
\usepackage{epstopdf}
\usepackage{mathtools}

\title[Ice Aggregate Contacts at the nm-Scale]{Ice Aggregate Contacts at the nm-Scale}
\author[G. Aumatell and G. Wurm]{Guillem Aumatell$^{1}$\thanks{E-mail:
guillem.aumatell@uni-due.de} and Gerhard Wurm$^{1}$\\
$^{1}$Fakult\"at f\"ur Physik, Universit\"at Duisburg-Essen, Lotharstr. 1, D-47057 Duisburg, Germany}

\begin{document}

\date{Accepted date. Received date; in original form date}

\pagerange{\pageref{firstpage}--\pageref{lastpage}} \pubyear{2002}

\maketitle

\label{firstpage}

\begin{abstract}

Aggregation of ice particles is a fundamental process in the interstellar medium as well as in planet formation. Dedicated to study the contact physics of nm-ice particles we developed a thermal gradient force microscope. This allows us and was used to measure pull-off forces with a resolution on the nN-scale and to measure rolling. Furthermore, based on a free probe, it also allows us to study twisting torques for the first time. The experiments show that torques required to twist are significantly larger than that macroscopic models scaled down to the nm-size range would predict. This implies that (ice) aggregates in astrophysical settings with small constituents are more robust against restructuring than previously thought. They likely grow as fractal aggregates to larger size before they restructure and during later compact growth they likely retain a higher porosity during further evolution towards planetesimals.

\end{abstract}

\begin{keywords}
solid state: volatiles - planets and satellites: formation - protoplanetary discs 
\end{keywords}

\section{Introduction}

Among the most abundant solid materials in protoplanetary discs are silicates and water ice. Inwards of the snowline no free ice is present and the size evolution of silicates dominates the formation of planetesimals and planet formation. This is a multi-step process. It starts by the aggregation of micrometer particles in sticking collisions as fractal growth \citep{Blum2000b, Kempf2000,  Paszun2009}. It continues with sticking collisions up to mm-size \citep{Dominik1997, Wurm_Blum1998, Blum_Wurm2000, Wada2011}. Depending on the contact properties of the monomers, fractal growth could continue to much large sizes \citep{Okuzumi2012}. The experimental measurements on ice as reported in this paper can be used to better quantify these transitions between the growth regimes.

The further evolution is more complex. Concentration of particles in pressure bumps, in stable eddies, by turbulence -- once thought to be an obstacle to concentration -- or by streaming instabilities might enhance the solid particle number density enough to lead to a gravitational collapse \citep{Goldreich1973, Safronov1967, Weidenschilling1993, Johansen2006, Dittrich2013, Chiang2010, Youdin2002}. There is little doubt though that collisions continue to be important processes. In fact the formation of planetesimals can also be built on sticking collisions. The current idea here is that a bouncing barrier at mm or cm-size exists which essentially prevents further growth of most of the particles \citep{Zsom2010, Kelling2011}. Further growth is then possible if some particles grow large by chance or are introduced otherwise \citep{Windmark2012a}. There are several possibilities to provide these large seeds for growth, either lucky conditions (small collision velocities), large constituent grains or the aggregation of some initial aggregates \citep{Kothe2010, Jankowski2012, Windmark2012b}. Seeds might also be drifting in from further outward, i.e. might grow around the snowline and lose the water while drifting inwards \citep{Saito2011, Sirono2011, Aumatell2011, Drazkowska2013}.

Once larger aggregates exist, collisions include fragmentation and growth \citep{Wurm2005, Teiser2011, Teiser2012, Kothe2010, Beitz2011, Meisner2012, Schaefer2007,Dove2012}. A summary of collisional outcomes relevant to protoplanetary discs has been given by \citet{Guettler2010}. 

The fundamental reason why dust particles are kept together after a collision is sticking forces acting at the contacts, like van der Waals forces, dipole interactions or surface tension. Depending on the strength of the contacts, growth is possible in collisions or not. If growth is possible the contact strength defines the structure of the growing aggregate, i.e. the transition from a hit-and-stick collision to compaction if the contact is weak enough to allow restructuring.

For spherical macroscopic particles, a theoretical treatment is possible which in agreement to the experiments mentioned above can be scaled down to micrometer dust particles \citep{Chokshi1993, Dominik1995, Dominik1996, Wada2007}. This will be quantified in later sections. However, it is not clear what the smallest size is to which these models can be extended and be used. Specific experiments related to rolling of microspheres seem to be in agreement with the mentioned theory \citep{Ding2007, Sumer2008}. If going to nm-scale, where the interaction of individual atoms or molecules becomes important, a down scaling might be questionable but experiments are rare (e.g. \cite{Asif1999}). 

The situation gets even more complex for water ice. As ice is rather volatile, sublimation and condensation have to be considered in protoplanetary disks \citep{Aumatell2011, Saito2011, Sirono2011, Ros2013}. It has also been suggested that collisions with energy enough to heat the contacts can lead to collisional fusion \citep{Wettlaufer2010}. Otherwise, in individual collisions again the sticking forces are important to decide what the outcome of the event is. Experiments with macroscopic ice samples have been carried out in the past \citep{Bridges1996, Higa1998, Arakawa2000, Heisselmann2010}. Experiments on the sticking of micrometer ice grains only began recently \citep{Gundlach2011}.

Ice particles are often considered along with dust particles but the contacts are stronger and provide additional aspects in contact physics (see below). In general the dynamical properties related to a contact can be divided into four parts, the break-up force needed to break up a contact, the rolling torques allowing two particles in contact to roll over each other, forces related to sliding if two particles slide over each other (but do not rotate), and the twisting around a given contact which from the physics is related to sliding (a circular sliding around a fixed point on a surface). Depending on the relation in strength between these processes which are tied to particle size and the structure of a given aggregate the different processes are more or less important, e.g. with respect to restructure an aggregate \citep{Dominik1997, Geretshauser2011, Seizinger2012, Kataoka2013}. \citet{Kataoka2013} e.g. found that depending on the porosity of an aggregate either rolling friction or twisting friction dominates energy loss.

Measurements of contact forces of micrometer dust are consistent with theory. Most are related to break-up and rolling \citep{Heim1999, Ding2007}. Experiments on twisting effects are very rare \citep{Sumer2008}. In lack of data for sliding but due to the need to match simulations and experiments e.g. \citet{Seizinger2012} assumed contacts to be stiffer than predicted by existing models. To our knowledge no data exists on torsion (twisting torques) and especially not on the nm-scale. A general problem with measuring torsion is that e.g. the cantilever of an AFM as used by \citet{Heim1999} is fixed and not free to rotate. We therefore developed a new method using a 'free probe' to measure torsion, rolling and breakup, so far for water ice contacts of nm-size as reported here.

This paper is structured as follows. We first describe the experimental idea to measure forces at the nN-scale in section \ref{albert}. We then give a specific realization used to study ice aggregate contacts in section \ref{josef}. In section \ref{schaf} we show our measurements of the pull-off forces, the twisting torques and the rolling torques. Section \ref{marina} summarizes the current theoretical ideas for large microscopic particles. In section \ref{wasserstoff} we discuss our data with respect to theory and estimate correction factors based on the experimental results. Section \ref{hundeleine} is dedicated to open issues of the experiment. Section \ref{wassersack} concludes this paper.

\section{A thermal gradient force microscope}

\label{albert}

State-of-the-art instruments to measure forces on the nN-scale are atomic force microscopes (AFM). They have been used in the past to measure the pull-off force on micrometer-size dust grains \citep{Heim1999}. The basic functionality used is the detection of the deflection of a light beam by a cantilever flexing under load. An AFM in contact mechanics has the advantage that well-defined forces pulling or compressing can be applied. However, it has the disadvantage that a sample is fixed in two points and is not free to rotate around a single contact. Therefore, twisting around
one contact cannot be measured easily.

Ideally, in order to study rotation around a contact, the probing 'tip' has to be capable of free rotation as well, e.g. following the motion of a particle being in contact with a surface or other particle. The 'tip' also has to have the ability to apply a force and -- in the case of torsion to be studied -- it also has to apply a torque.
 
\begin{figure}
 \centering
\includegraphics[width=1.0\columnwidth]{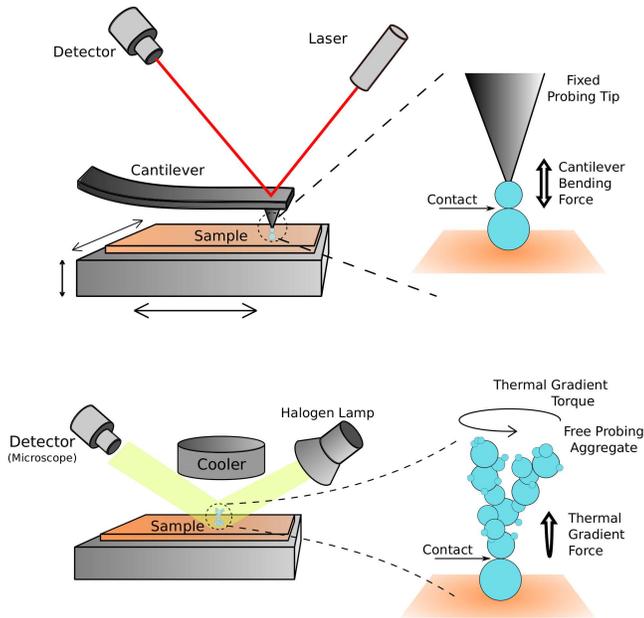}
 \caption{In comparison to an AFM our {\it thermal gradient force microscope} uses a free floating probe (aggregate) to measure pull-off force, rolling torque and twisting torque.}
 \label{probe1}
\end{figure}

Keeping the analogy to the AFM for a while, instead of the cantilever of sub-mm size, we therefore need a free floating probe on sub-mm scale which can apply forces and torques, and the motion of which can be detected. In general, a free floating probe has to be attached to the measuring contact and an external field has to provide a force pulling the particle and providing the torque. As a free floating probe is not fixed like a cantilever, it does not necessarily flex under load in a well-defined way. The easiest way to track the motion then is microscopic observation. The torque is  given by the angular acceleration around an axis if the moment of inertia of the probe (and sample) is known. The pull-off force of a contact can be determined by observation of the linear acceleration of the probe if the contact breaks and if the total mass is known. These ideas are sketched in Fig. \ref{probe1}.

As for the cantilever of the AFM where the elastic bending is measured as mechanical quantity we have an equally simple mechanical analysis here.
For the pull-off force $F_{\text{c}}$ with a probe mass $m_{\text{p}}$ it is
\begin{equation}
  F_{\text{c}} = m_{\text{p}}\:a
  \label{secondnewton}
\end{equation}
The acceleration $a$ is taken from the observation of the centre of mass of the probe after the contact breaks. For the torques the usual equations apply as well.
\begin{equation}
M = I\:\alpha,
\end{equation}
where $\alpha$ is the angular acceleration, and $M$ and $I$ are the torque and moment of inertia with respect to the rotational axis under consideration, either around the vertical in the case of torsion or around a horizontal axis in the case of rolling (see below). In the current version of the setup, the detection of motion requires a microscopic motion of the probe. We therefore only detect rolling or torsion if the elastic limit is exceeded. The break-up force can only be measured once as the sample cannot be recovered after the measurement.
Therefore, with a free floating probe, three measurements of a contact are possible in principle:
\begin{itemize}
\item rolling torque, 
\item twisting torque,
\item pull off force (final measurement).
\end{itemize}
Experiments with free probes require a pull on the probe. Otherwise, the probe will just follow gravity downwards and settle itself on the surface. This also implies that torsion can only be measured if the contact does not completely break once the torque is sufficient to initiate rotation around the vertical axis. This is supposed to be the case at least for water ice and metals \citep{Dominik1997}.

\subsection{A thermal gradient probe}

As external field exploited here, we use the thermophoretic force acting on a particle in a low-pressure atmosphere and in a temperature gradient field (Fig. \ref{probe1}). The thermophoretic force for a spherical particle at low pressure is given as \citep{zheng2002the} 
\begin{equation}
F_{\text{th}}=-f\frac{a^2\kappa _{\text{g}}}{\sqrt{2k_{\text{B}} T_0/m}}\nabla T,
 \label{thermoforce}
\end{equation}

where $a$ is the particle diameter, $\kappa_{\text{g}}$ the thermal conductivity of the ambient gas, $m$ the mass of a gas molecule, $k_{\text{B}}$ the Boltzmann constant and $T_0$ the average temperature at the particle, and $f$ is a dimensionless parameter depending on the gas pressure. For a 1 $\mu$m particle surrounded by air ($\kappa_{\text{g}}\approx$ 0.01 $\text{Wm}^{-1}$K, $m\approx 4.8\times 10^{-26}$ Kg) at a pressure of 0.5 mbar, temperature of 200 K and a temperature gradient of 4100 $\text{Km}^{-1}$, the thermophoretic force is $\approx 2.3\times 10^{-13}$ N.

As thermophoresis in a very open structured aggregate pulls on each individual constituent,  different size aggregates are related to different forces, the larger the aggregate the larger the force. Due to non-sphericity there are some variations in the force, which cannot be adjusted to high degree but show some random component in the strength. However, by adjusting the aggregate size and the ambient pressure the force can be varied in principle.

In the specific case here, we use microscopic ice aggregates as probes. If the probe is of the same material as the sample to study it actually provides the contacts. The ice aggregates used are non-symmetric. This means that thermophoretic forces also show components perpendicular to the direction of the temperature gradient. This always leads to a small random twisting torque around the vertical \citep[direction of temperature gradient;][]{vanEymeren2012}. It is this torque, which allows a measurement of the strength of a contact with respect to twisting. Besides contact physics (the focus in this work), the observations also provide means to study the thermophoretic forces and torques, and from rotation frequencies in equilibrium the rotational gas-grain coupling times. We will sketch this below.

\subsection{Experimental setup}
\label{josef}

\begin{figure}
 \centering
 \includegraphics[width=.47\textwidth]{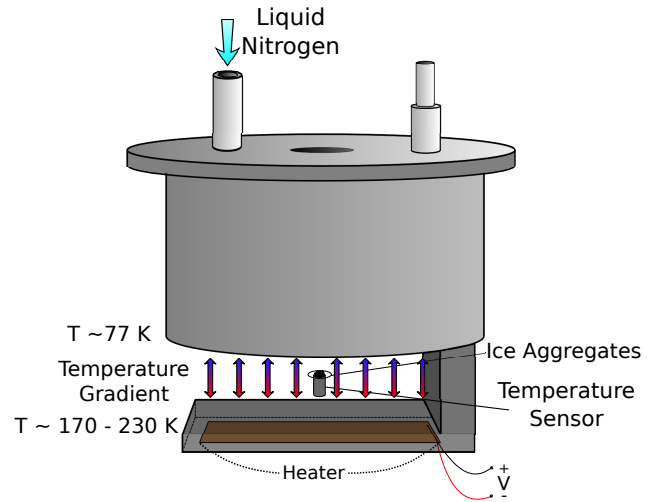}
 \caption{Thermal gradient generation and sample placement}
 \label{setup1}
\end{figure}

\begin{figure}
 \centering
 \includegraphics[width=.47\textwidth]{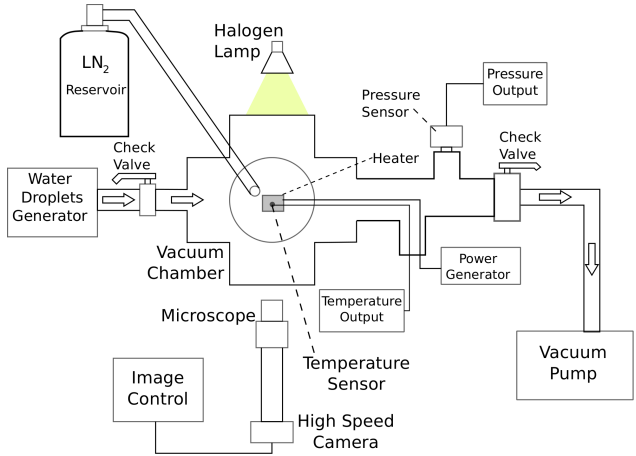}
 \caption{Overview of the principle experiment components.}
 \label{setup2}
\end{figure}

The main components of the specific setup can be seen in Fig. \ref{setup1} and Fig. \ref{setup2}. At 4 cm below a liquid nitrogen reservoir, there is a horizontally oriented copper plate attached to the bottom of the reservoir. On the bottom side of this plate, a heating foil is attached in order to create a temperature gradient above the plate.
Ice aggregates are placed on a substrate within the temperature gradient field and are imaged by a long distance microscope with working distance of 18 cm. We took bright field images with particles being silhouette in front of a light source. Frame rates varied between 2 frames second$^{-1}$ for the sublimation rate studies to high speed observations at 800 frames second$^{-1}$ for observation of rotation and break up.

\subsection{Ice aggregate generation}

\begin{figure}
 \centering
 \includegraphics[width=.47\textwidth]{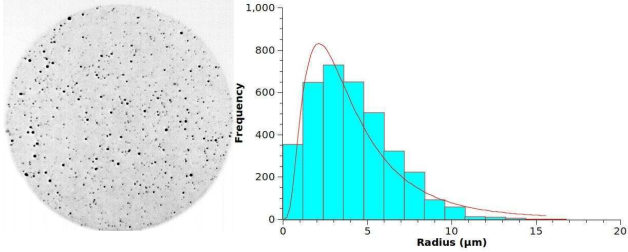}
 \caption{Left: example of liquid water droplets used to form ice aggregates. Right: size distribution of water droplets.}
 \label{droplets}
\end{figure}

Water ice aggregates are formed from frozen water droplets. The droplets are generated with a vapourizer. Fig. \ref{droplets} shows an image of the droplets in-flight after they  travelled a tube of 10 cm length and passed between two glass plates. The corresponding size distribution is also plotted showing a peak at 2.1 $\rm \mu m$. The ice particles are transported to the temperature gradient region below the liquid nitrogen reservoir. At temperatures below 180 K, the water particles freeze. 

\begin{figure}
 \centering
 \includegraphics[width=.47\textwidth]{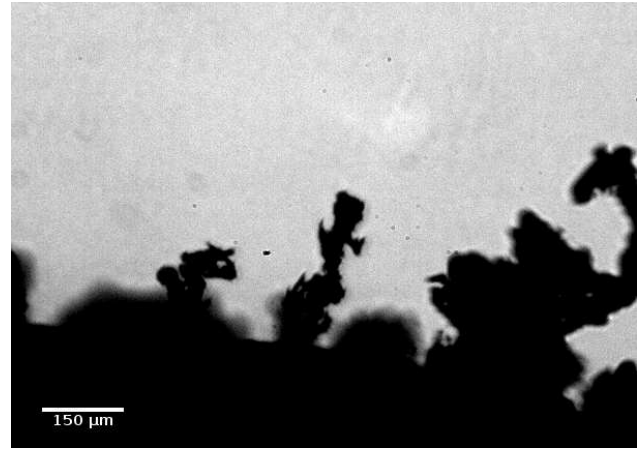}
 \caption{A layer of ice aggregates forms on top of the temperature sensor.}
 \label{aggregates}
\end{figure}

At normal atmospheric pressure, turbulence within the experiment chamber leads to collisions between the ice particles and growth of aggregates. As substrate a 2 mm cylinder-shaped temperature sensor is placed within the low-temperature region and is used as a target where ice aggregates attach themselves and grow. After a few minutes of particle injection, the sensor surface is covered by a large number of ice aggregates. The thickness of this aggregate layer depends on the deposition time, with sizes ranging between some $\mu$m to a few mm (Fig. \ref{aggregates}). As we use a temperature sensor as substrate for the ice aggregates, it is possible to determine the temperature of the ice aggregates at any moment. 
Once the aggregates are formed, the chamber is evacuated to a pressure between 0.1 and 1 mbar (so far), and thermophoretic forces act on the particles.  

\subsection{Sublimation}
\label{subsec:sublimation}

At this pressure and at temperatures of about 180 K, the sublimation rate is too low to have visible effects at $\mu$m size at time-scales of seconds. 
At a higher temperature of about 200 K. the ice samples slowly start to sublimate. In the experiments reported here, we use the sublimation to reduce the contact size but keep the temperature gradient constant. The ice experiments therefore make use of sublimation which decreases the contacting particle size and contact area until the applied torques and pull-off forces are stronger than the contacts can compensate, eventually. The onset of twisting and the break-up occur at different times. Due to sublimation this relates to different particle sizes. We calculated this difference based on the measured \textbf{and calculated} sublimation rates. The sublimation rate \textbf{for large grains} was measured as follows.
The initial particles are homogeneous spheres formed by water droplets. \citet{Saito2011} showed that the rate at which a sphere's radius $R$ decreases is given as

\begin{equation}
 \frac{dR}{dt}=-\frac{P_{\text{ev}}(T)-P_{\text{H}_2 \text{O}}}{\rho_{\text{H}_2 \text{O}}}\sqrt{\frac{m_{\text{H}_2 \text{O}}}{2\pi k_{\text{B}} T}}
 \label{sublimation_rate}
\end{equation} 
where $P_{\text{H}_2 \text{O}}$ is the partial H$_2$O gas pressure, $\rho_{\text{H}_2 \text{O}}$ is the water ice density, $m_{\text{H}_2 \text{O}}$ is the mass of a water molecule, $T$ the ice temperature and $P_{\text{ev}}$ the equilibrium vapour pressure that depends on the temperature as \citep{Yamamoto1983}
\begin{equation}
\begin {split}
 \log_{10} P(T)_{ev} & =-2445.5646/T + 8.2312log_{10}T \\
 & -0.01677006T+1.20514\times 10^{-5}T^2 \\
 & -3.63227.
 \end{split}
  \label{Pev}
\end{equation}

As seen in equation \ref{sublimation_rate}, the shrinking in radius difference per time does not depend on the absolute radius of the droplet. Under the experimental conditions, shrinking of large individual water ice droplets with a well-defined diameter is tracked. Measuring the time $\Delta t$ required for the droplet to shrink from a radius $R_1$ to a radius $R_2<R_1$ immediately gives the sublimation rate. We find 
\begin{equation}
dR/dt = 0.032\pm 0.014\: \mu \text{m\:s}^{-1} 
\end{equation}
at 203 K and a total pressure within the chamber of 0.66 mbar. From equation \ref{Pev} we calculate $P_{\text{ev}}$ to 0.025 mbar. Using equation \ref{sublimation_rate} and in view of the small sublimation rate observed, the partial water pressure $P_{\text{H}_2\text{O}}$ is comparable to $P_{ev}$. This implies almost saturated conditions.

For the nm-particle which sets the strength of the aggregate sublimation is important. Equation \ref{Pev} is only valid for sub-$\mu$m size or for larger particles, but for smaller radii, the vapour pressure and therefore the sublimation rate show a strong dependence on the surface curvature $K$. The equilibrium vapour pressure $P_s$ is \citep{Sirono2011b}

\begin{equation}
\ln \left(\frac{P_s}{P_{ev}}\right)=K\frac{\gamma v}{k_{\text{B}} T}
\label{P_s}
\end{equation}
where $\gamma$ is the surface energy of ice, $v$ the molecular volume, $k_{\text{B}}$ the Boltzmann constant and $T$ the temperature.

The vapor pressure $P_{\text{ev}}$ has to be substituted by $P_s$ 
in equation \ref{sublimation_rate}. If we further assume that the curvature of the nm-contacting grains is $K \sim 1/R$ we obtain as sublimation rate

\begin{equation}
 \frac{dR}{dt}=-\frac{\sqrt{\frac{m_{\text{H}_2 \text{O}}}{2\pi k_{\text{B}} T}}}{\rho_{\text{H}_2 \text{O}}}\left(P_{\text{ev}}\text{e}^{\left(A/R\right)}-P_{\text{H}_2 \text{O}} \right)
 \label{sublimation_rate2}
\end{equation}

with $A = \frac{\gamma v}{k_{\text{B}} T}$.

Equation \ref{sublimation_rate2} was used to calculate the particle sizes for the time twisting sets in. During the short sublimation times of $\sim$ 0.1 s (see below) between start of rotations and break-up, the much larger aggregate probe attached to this particle, which consists of micrometer particles, only changes insignificantly. Therefore, thermal gradient forces and torques do not change during this short time. This is also consistent with the observation that the aggregate does not change on the microscope images during the observation times of an image sequence of 0.5 s at a frame rate of 800 frames seconds$^{-1}$.

\section{Breakup Force}
\label{schaf}

\begin{figure}
 \centering
 \includegraphics[width=.47\textwidth]{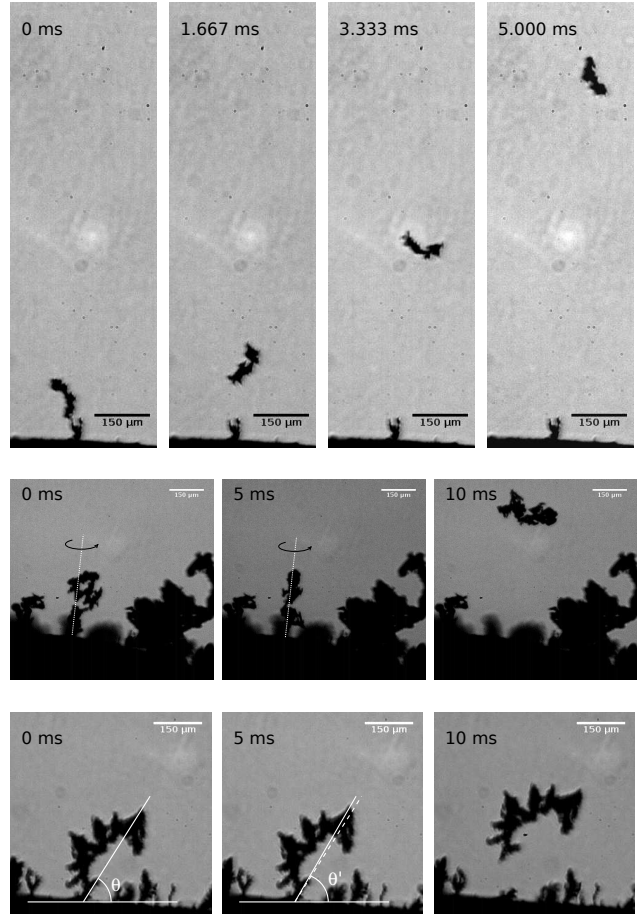}
 \caption{Time sequences of aggregates with respect to break-up (top), twisting (middle) and rolling oscillations (bottom). For each sequence, the time passing in milliseconds is shown.}
 \label{sequence}
\end{figure}

As the temperature gradient force is directed from the cold to the warm side, the net direction of the thermophoretic force is upwards. Break-up of contacts and ejection of the 'probe' aggregates are observable for all aggregates. Break-up will take place sooner or later due to sublimation and depending on the weakest contact with contact area $a_0$. Some of these aggregates do not present any particular motion before they break up, move upwards and disappear from the field of view (Fig \ref{sequence} (top)). On the other hand, many of them present oscillating (Fig. \ref{sequence}, bottom) or twisting motions (Fig \ref{sequence}, center) (see below). 

As mentioned above, all break-up forces are calculated applying Newton's second law $F=ma$ to the motion of an aggregate after break-up. Accelerations were determined from the displacement of the aggregates (visible) centre of mass on all frames while imaged and the timing between frames. This force measurement requires the determination of the acceleration and the aggregate mass.

\subsection{Determination of the aggregate mass}

\label{subsec:mass}

Sequences of aggregate images were taken when they were rotating around their vertical axis and a view at different angles was possible. As aggregates often showed compact regions which could not be resolved in any perspective we did not attempt to do a full 3D reconstruction based on all pixels on all perspectives. Instead, the particle masses were extrapolated from an approximated 3D structure. For this, the aggregate is approximated by a projected area $A$ multiplied by a thickness $x$.
 
\begin{figure}
 \centering
 \includegraphics[width=.47\textwidth]{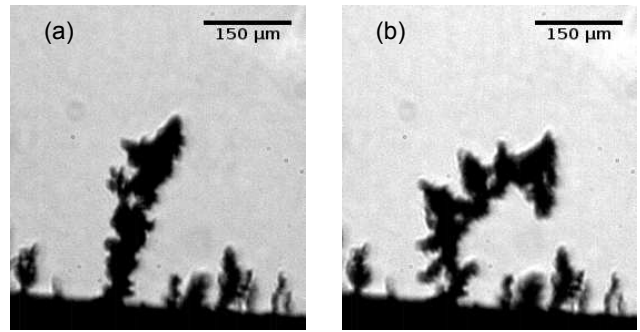}
 \caption{Two images from a rotation sequence used for mass determination; 
 (a) $90^o$ rotated for thickness determination, (b) largest cross-section,}
 \label{massdet}
\end{figure}

Observing a full $360^o$ rotation sequence, two frames were chosen. One with the maximal cross-section $A$ and a second image where the particle rotated by $90^o$ and showed a cross-section  $A'$. The thickness $x$ is then calculated from the ratio $x=A'/L$ where $L$ is the aggregate's length along the rotation axis as e.g. shown in fig. \ref{massdet}.

The mass estimated this way does not account for hollow parts invisible and therefore is an upper limit. It can be corrected by assuming a filling factor smaller than 1. A value for the filling factor close to 1 is unlikely as this would imply a rigid solid aggregate. A very low value is not adequate to assume as might e.g. result from a pure fractal structure as the aggregation process in which the ice structure forms includes sublimation, sintering, and condensation. As an exact filling factor is not possible to obtain with the used equipment, we opted for a filling factor of 0.4 and assumed a factor 2 as the uncertainty which only rules out the extremes. This also constrains upper and lower values for its derivative variables below.

\subsection{Thermophoresis for aggregates}

The aggregates are grown through collisions and are highly porous  though this is not considered in further detail here. In any case, as a first approximation, gas can flow freely around the constituents. This suggests that the thermophoretic force is proportional to the number of constituents (if monodisperse) or total mass under otherwise same conditions (temperature gradient and pressure). 

\begin{figure}
 \centering
 \includegraphics[width=.47\textwidth]{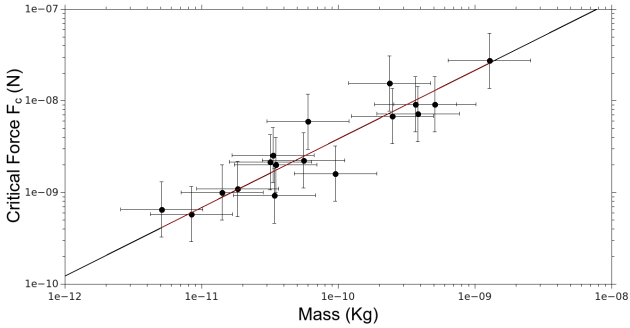}
 \caption{Thermophoretic force depending on aggregate mass. The line indicates
 a linear dependence.}
 \label{forceovermass}
\end{figure}

In Fig. \ref{forceovermass} the thermophoretic force at break-up is seen in dependence of the aggregate mass. The forces determined by the accelerating particles have been corrected for gravity, i.e. subtracting $m g$. As a general trend, the data are consistent with a linear mass dependence plotted as a solid line. A detailed study of thermophoretic forces on aggregates is possible with the given setup but this is beyond the focus of this work. This might be exploited in future applications of this technique. Here, it is a consistency check that thermophoretic forces are responsible. Important is that, in principle, larger aggregates provide larger forces and can probe larger contact forces. 

Giving the mass and the break-up force, we can give first estimates of the reduced radius and the contact area of the contact breaking. This requires a theory for the contact force. Noting that this does not exist yet for the nm-scale as outlined below, we use the equation which proved useful on the micro-scale \citep{Johnson1971, Dominik1997, Heim1999}
\begin{equation}
 F_{\text{c}}=3\pi\gamma R,
 \label{critforce}
\end{equation} 
where $\gamma$ is the surface energy. For ice, literature values for $\gamma$ vary between 0.1 $\rm Jm^{-2}$ and 0.37 $\rm J/m^2$ \citep{Dominik1997, Wada2007, Gundlach2011, Kataoka2013}. In Table \ref{mainvalues} we used $\gamma = 0.37$ $\rm Jm^{-2}$. For the contact area radius we used equation \ref{a} as detailed below.

\begin{table*}[htbp]
\caption{Parameters for ice contacts at breakup: break-up force $F_{\text{c}}$, reduced radius $R$ and contact radius $a_0$. Uncertainties are estimated to a factor of 2 and result from the mass uncertainty. We assume a surface energy of $\gamma = 0.37$ $\rm Jm^{-2}$ and a filling factor of 0.4.}
\centering
\begin{tabular}{|l|c|c|c|c|}
\hline
\multicolumn{1}{l|}{Id} & \multicolumn{1}{l|}{Mass} & \multicolumn{1}{l|}{$F_c$} & \multicolumn{1}{l|}{$R$} & \multicolumn{1}{l|}{$a_0$} \\
   & (pg) & (nN)  & (nm) & (nm)\\ \hline\hline
1 & 508 & 9.18 & 2.63 & 2.74 \\ \hline
2 & 366 & 9.16 & 2.63 & 2.74 \\ \hline
3 & 237 & 15.5 & 4.45 & 3.89 \\ \hline
4 & 31.8 & 2.15 & 0.62 & 1.00 \\ \hline
5 & 34.8 & 2.00 & 0.56 & 1.00 \\ \hline
6 & 382 & 7.16 & 2.05 & 2.33 \\ \hline
7 & 1272 & 27.4 & 7.86 & 5.69 \\ \hline
8 & 60.0 & 5.93 & 1.70 & 2.05 \\ \hline
9 & 33.2 & 2.55 & 0.73 & 1.17 \\ \hline
10 & 95.2 & 1.60 & 0.46 & 0.86 \\ \hline
11 & 34.1 & 0.93 & 0.25 & 0.59 \\ \hline
12 & 8.40 & 0.58 & 0.17 & 0.44 \\ \hline
13 & 55.6 & 2.23 & 0.64 & 1.07 \\ \hline
14 & 248 & 6.81 & 2.00 & 2.25 \\ \hline
15 & 14.1 & 1.00 & 0.30 & 0.64 \\ \hline
16 & 5.11 & 0.65 & 0.19 & 0.47 \\ \hline
17 & 18.3 & 1.09 & 0.31 & 0.66 \\ \hline
\end{tabular}
\label{mainvalues}
\end{table*}

\section{Contact physics for large grains}

\label{marina}

There are a number of theoretical works on the forces and torques acting when two particles get in contact. \citet{Johnson1971} and \citet{Dominik1997} calculated the strength of a contact or the break-up force necessary to separate two particles again. \citet{Dominik1995} and \citet{Dominik1996} studied different aspects, rolling, sliding and twisting. This is used in \citet{Dominik1997} to calculate the behaviour of particle aggregates. \citet{Wada2007} used a different approach but got similar results as in \citet{Dominik1997}. Experiments by \citet{Heim1999}, \citet{Wurm_Blum1998} and \citet{Poppe2000a} showed that these descriptions are adequate for micrometer silicate particles. The following equations for the different forces and torques are taken from \citet{Dominik1997}. The force at which a contact breaks was used above and is given as equation \ref{critforce}. The radius of the contact area depends on the force $F$ applied and is 
\begin{equation}
a = \left(\frac{3R}{4E^*}\left(F+6\pi\gamma R+\sqrt{(6\pi\gamma R)^2+12\pi\gamma RF}\right)\right)^{1/3}
\label{a}
\end{equation} 
with the reduced module of elasticity $E^{\star}$ which is $0.5E$ in the case of only one material. The equilibrium radius (no external force) is
\begin{equation}
 a_0=\left(\frac{9\pi\gamma R^2}{E^{\star}}\right) ^{1/3} 
 \label{a0}
\end{equation} 

\begin{figure}
 \centering
 \includegraphics[width=.47\textwidth]{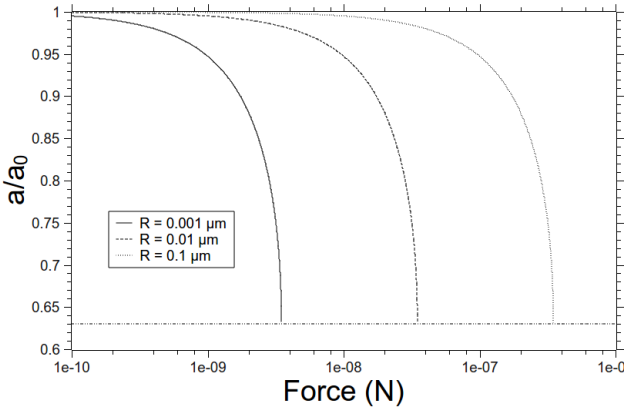}
 \caption{The ratio between contact area $a$ under load and the equilibrium contact area $a_0$ depending on the pulling force. The three curves correspond to different reduced radii $R$ = 0.001, 0.01 and 0.1 $\mu$m.}
 \label{aa0}
\end{figure}

If the applied force $F$ pulls on one of the two monomers, the contact area will decrease as shown in equation \ref{a} until the contact breaks. The decrease in $a$ is shown in Fig \ref{aa0} as a normalized parameter $a / a_0$. The contact area will break once the pulling force is equal to the critical force $F_{\text{c}}$. Substituting $F_{\text{c}}$ for $F$ in equation \ref{a} we obtain $a/a_0 = (1/4)^{1/3} = 0.63$ independently of the reduced radius value $R$. This is the smallest area before the breakup takes place.

The torque at or beyond the threshold for non-elastic rolling is

\begin{equation}
 M_r=4F_{\text{c}}\left( \frac{a}{a_0}\right) ^{3/2} \xi_{\text{c}}
\label{Mr}
\end{equation} 
with the critical distance $\xi_{\text{c}}$ being on the order of 0.1 nm, but this parameter is not determined by the theoretical model. The torque for non-elastic twisting is
\begin{equation}
 M_t=\frac{G^{\star}a^3}{3\pi} + \frac{\pi}{3}F_{\text{c}}a_0\left( \frac{3}{4}\left( \frac{a}{a_0}\right) ^4 - \left( \frac{a}{a_0}\right)^{5/2}\right) - \frac{2}{9}\pi a^3p_{\text{c}}
 \label{Mt}
\end{equation}
 
where $p_{\text{c}}$ is given as

\begin{equation}
 p_{\text{c}}=\frac{2.67b^3}{\pi\sigma^3}G^{\star} - \frac{24.72b^4}{\pi\sigma^5}\gamma
 \label{pc}
\end{equation} 
and $G^{\star}$ is the reduced shear modulus which is $0.5G$ considering the same material (water ice) for contacting particles. Further constants are $b$ as inter-atomic distance in the grain material and $\sqrt[6]{2\sigma}$ as equilibrium distance in the pair-potential model between atoms of the two contacting surfaces.

In total these equations contain a few material parameters $\gamma, b, E, G, \xi_{\text{c}} and \sigma$. The constants $\gamma, E$, and $G$ are macroscopic quantities and constrained to some level. $E$ is $7\times 10^9 \text{N\:m}^{-2}$, and the shear modulus is $G = 2.8\times 10^9 \text{N\:m}^{-2}$ \citep{Anderson1981}. These values might vary for different ice phases and temperatures but we consider them to be given for the moment. The constants $b = 0.336$ nm and $\sigma = 0.336$ nm refer to atomic scales \citep{Dominik1997}.

According to \cite{Dominik1997} equation \ref{Mt} is valid for ice (and e.g. iron). For silicate particles, the second and third terms do not exist. The basic quantity, unknown but relating model and experiment, is the reduced radius of the contacting particles. However, the break-up force is linearly connected to the reduced radius with $\gamma$ being the only material parameter that has to be known. We use this to (at least formally) determine a reduced size from our measurements of the break-up force. 

\section{Rotations}

\label{wasserstoff}

\subsection{Rolling}

\begin{figure}
 \centering
 \includegraphics[width=.47\textwidth]{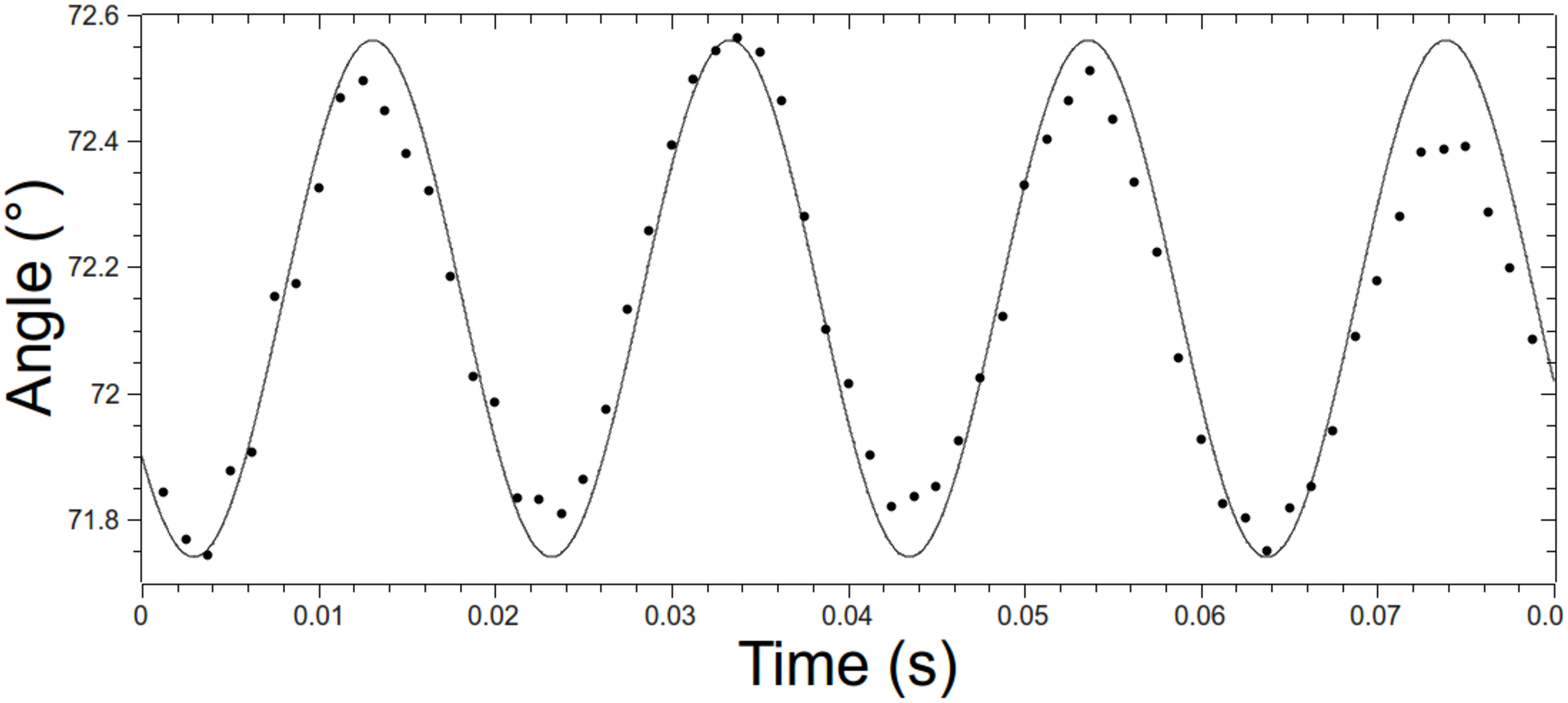}
 \caption{Oscillation (rotation at a contact) of an aggregate. The solid line corresponds to a harmonic function fitted to the data.}
 \label{golem_oscillation}
\end{figure}

Some aggregates show an oscillation along a perpendicular direction to the lifting force resembling similarities to an upside down pendulum. The amplitude of these oscillations is small compared to the size of the aggregate (between 1$^{\circ}$ and 10$^{\circ}$), but measurable. The oscillations are not bending throughout the whole aggregate. The aggregate remains rigid and the motion can be traced to a rotation along one point. Therefore, the oscillation is related to the physics of rolling along the contact. An example of an aggregate showing this rotation is shown in Fig. \ref{sequence} and the data for the oscillation are given in Fig. \ref{golem_oscillation}.

\begin{figure}
 \centering
 \includegraphics[width=.47\textwidth]{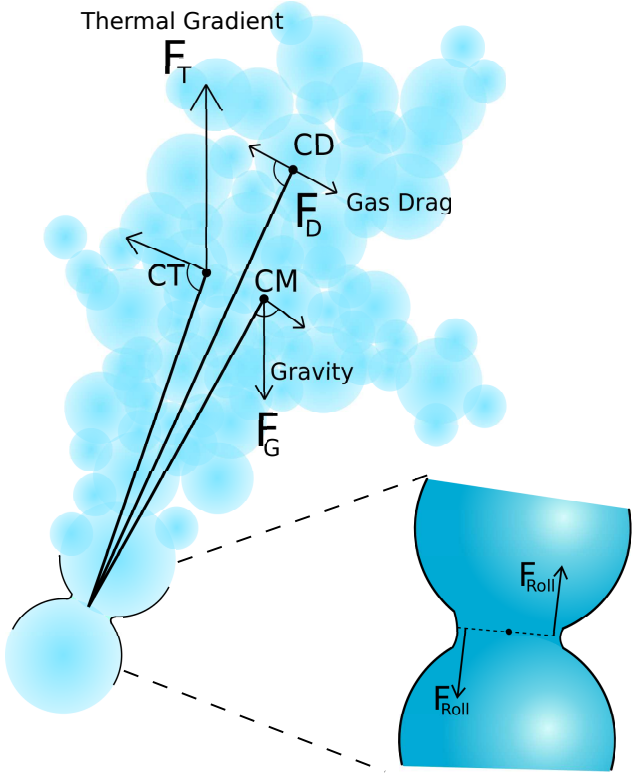}
 \caption{Sketch of all rolling torques acting on an aggregate.}
 \label{rollingsketch}
\end{figure}

These oscillations can well be approximated by harmonic functions. A constant harmonic oscillation in angle requires (1) a restoring torque linear with displacement, (2) an exciting torque and (3) a damping torque. The exciting torque results from the substrate or temperatures sensor to which frequencies of about 100 Hz couple from the laboratory environment. For the restoring torques and damping different possibilities exist a priori. This is visualized in Fig. \ref{rollingsketch} as follows.

\begin{itemize}
\item Restoring torque: elastic bending at the contact, resistance to rolling,
\item Restoring torque: component of the thermophoretic lifting force if
the aggregate is displaced. Motion due to this torque kind of compares to a simple physical pendulum under gravity. Only the direction of the force (upwards instead of downwards) is inverted.
\item Gravitational torque: for an aggregate with resistance free contact, thermophoretic torque and gravitational torque need to cancel out in the equilibrium position, but displaced aggregates can show varying gravitational torques. 
\item Damping: resistance within the contact area once displacements are too large for elastic torques. 
\item Damping: gas drag or friction with the remaining gas due to the motion of the aggregate oscillating.
\end{itemize}

We can measure the restoring torque using the oscillation. As damping and excitation balance each other on average the restoring torque is determined by 
\begin{equation}
 M = I \alpha,
 \label{torque}
\end{equation} 
where $I$ is the  moment of inertia  and $\alpha$ its angular acceleration. The  moment of inertia is given as

\begin{equation}
 I = \sum\limits_i = m_i r_i^2,
\label{inertia}
\end{equation} 

where the index $i$ corresponds to each pixel of the aggregate image, $r_i$ is the distance between the $i$-th pixel and the point where the aggregate is sustained. The mass per pixel $m_i$ is chosen such that each pixel has the same mass and the total mass adds up to the aggregate mass given above. The mass associated with each pixel is equal to the water ice density ($\sim 0.92$ $\text{g\:cm}^{-3}$ at 200 K) times its area (3.533 $\pm$ 0.011 $\mu$m$^2$) multiplied by the thickness $x$ defined in Section \ref{subsec:mass}. As done before for the determination of the whole aggregate mass, we assume a filling factor of 0.4 with an associated factor of 2 uncertainty per pixel mass.

Torques related to rolling for a number of aggregates measured are listed in Table \ref{rolling} for the maximum elongation. $M_r$(osc) is the torque measured by the oscillations. $M_r$(therm) is based on the thermophoretic part of the break-up forces and assumed to act at the centre of mass (pendulum). The theoretical torques are taken from the model by \citet{Dominik1995} (see below). They are based on the size of the contact at break-up.

\begin{table*}[htbp]
\caption{Rolling torques as calculated by the model for the given contact $M_r$ (theor) for $\xi$ = 0.1 nm, as measured by the excited oscillations $M_r$ (osc), as determined from damped oscillations $M_r$ (damp), as estimated by the critical force $F_{\text{c}}$ times the projected distance between the centre of mass and the contact point at maximum elongation and the ratios between different torques.}
\centering
\begin{tabular}{|c|c|c|c|c|c|c|c|}
\hline
Aggregate & \multicolumn{1}{l|}{$M_r$ (theor) N$\cdot$m} & \multicolumn{1}{l|}{$M_r$ (osc) N$\cdot$m} & $M_r$ (damp) N$\cdot$m & \multicolumn{1}{l|}{$M_r$ (therm) N$\cdot$m} & \multicolumn{1}{l|}{$\frac{M_r (osc)}{M_r (theor)}$} & \multicolumn{1}{l|}{$\frac{M_r (damp)}{M_r (theor)}$} & \multicolumn{1}{l|}{$\frac{M_r (therm)}{M_r (osc)}$}\\ \hline
1 & 1.84E-018 & 1.55E-014 & -& 1.34E-014 & 8420 &-& 0.86 \\ \hline
2  & 1.83E-018 & 1.96E-014 & \multicolumn{1}{r|}{4.31E-015} & 6.46E-014 & 10700 & 2350 & 3.29 \\ \hline
6 & 6.83E-019 & 2.20E-013 & - & - & 322000 & - & - \\ \hline
7 & 5.48E-018 & 3.37E-014 & - & 4.70E-013 & 6150 & - & 13,95\\ \hline
13 & 4.47E-019 & 2.51E-015 & \multicolumn{1}{r|}{5.72E-016} & 4.92E-015 & 5610 & 1280 & 1.96 \\ \hline
14 & 1.36E-018 & 2.09E-015 & - & 6.93E-015 & 1530 & - & 5.09 \\ \hline
\end{tabular}
\label{rolling}
\end{table*}

A first hint that the contact is not providing the restoring torque is that the oscillation does not change with time and sublimation does not affect the torques.

Quantitatively, the measured torques are also way larger than the theoretical expectations for the contact torques. However, they are on the same order as the thermophoretic torques. This suggests that the contact contributes little to the restoring torque and thermophoresis (and gravity) dominate the restoring torques. The measurements are therefore an upper limit on the rolling torques which is consistent to the estimate of the much smaller rolling torques at least as deduced from existing models (see below). 

\begin{figure}
 \centering
 \includegraphics[width=.47\textwidth]{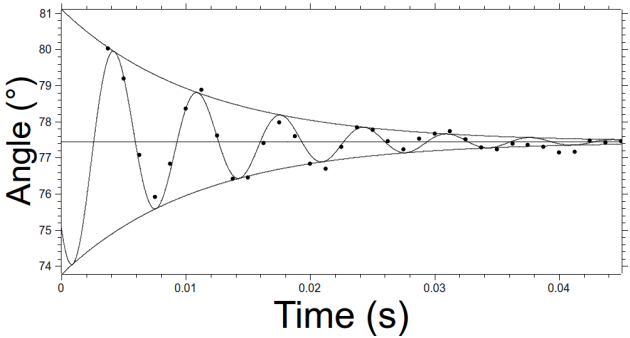}
 \caption{Damped oscillation (rotation at a contact) of aggregate 13. The solid line corresponds to an exponentially decaying harmonic function fitted to the data.}
 \label{Oscillations2}
\end{figure}

Further information on rolling can be deduced from oscillating aggregates where excitation occurs as a one-time event and where the amplitude then decays. We found two examples of such events. They were observed after some rearrangement of the aggregate, i.e. a second contact might have broken which allows rotation around a small contact left. Such an oscillation is seen in Fig. \ref{Oscillations2}.

In the damped case without continuous excitation the oscillation can be used to deduce the damping strength. This might either be dominated by gas drag or by the friction of the contact if the elastic limit is exceeded. As a friction force related to the contact is always of the same strength while the gas drag depends on the velocity there are two slightly different equations of motion for the damped oscillation. For the gas drag it is

\begin{equation}
 I\frac{\mbox{d}^2\theta}{\mbox{d}t^2}+\beta\frac{\mbox{d}\theta}{\mbox{d}t}+(F_{\text{T}}-F_{\text{G}}) l\theta = 0,
 \label{gasdrag}
\end{equation} 
where $I$ is the moment of inertia, $\beta$ is the damping constant, $F_{\text{T}}$ the thermophoretic force, $F_{\text{G}}$ the gravity and $l$ the distance from the contact point to the centre of mass. We assume $F_T$ and $F_G$ to act at the same point. Upper direction is chosen as positive. The general solution is

\begin{equation}
 \theta (t)= \theta_0 \mbox{e}^{-\beta t/2I} \cos(\omega_0 t+\varphi),
 \label{gasdragsol}
\end{equation} 

where $\omega_0 \equiv \left[ (F_{\text{T}}-F_{\text{G}})/I - (\beta/2I)^2\right]^{1/2}$ is the angular frequency, $\theta_0$ is the oscillation's amplitude and $\varphi$ the initial phase of the motion.
For the contact friction it is
\begin{equation}
 I\frac{\mbox{d}^2\theta}{\mbox{d}t^2} \pm M_r +(F_{\text{T}}-F_{\text{G}}) l\theta = 0.
 \label{friction}
\end{equation}
 
Here, $M_r$ is the contact friction torque which changes its sign each semi-oscillation, depending the direction on which the aggregate  moves. The solution can be written stepwise as \citep{Zonetti, Marchewka}

\begin{equation}
 \theta (t)=\theta^{\text{max}} \cos(w_0t+\varphi)+C,
 \label{frictionsol}
\end{equation}
where $\omega_0 = \left[ (F_{\text{T}}-F_{\text{G}})l/I\right]^{1/2}$, $C = M_r/I \omega_0^2$ and 
\begin{equation}
 \theta_n^{\text{max}} = \theta_0^{\text{max}} -2nC.
 \label{damping}
\end{equation}
For $n=0$, the aggregate is at its starting oscillation point, for $n=1$ it is oscillating in the other direction, etc. This process continues until it finally remains motionless. Using the identity defined in equation \ref{damping}, the equation \ref{frictionsol} may be written as \citep{Marchewka}
\begin{equation}
 \theta (t)=(\theta_0-2nC) \cos(w_0t)+(-1)^nC.
 \label{frictionsol2}
\end{equation}

In contrast to the case before, the amplitude does not decay exponentially but linear. This fact can be used to distinguish between contact friction and gas drag. A linear decrease is not consistent with the observed oscillation down to small amplitudes, and gas drag should be the dominating damping. This allows us to establish a maximum value for the effective rolling friction at the contact $M_r$, which will be lower than the total $M_r$(damp). Making use of equation \ref{frictionsol} for the data corresponding to aggregate 13 (Fig. \ref{Oscillations2}) and assuming that all the torque takes place at the contact point, we obtain the value for $C = F_rl/I\omega_0^2 = 8.05 \times 10^{-3}$ rad, and therefore the corresponding torque shown in Table \ref{rolling}. This sets an upper limit to the rolling torque due to contact friction as shown in the fourth column of Table \ref{rolling} as $M_r$ (damp). While this is closer to the theoretical value it is still a factor of 1000 larger. We note though that we currently only have two cases of this damped motion. As damping strongly indicates to be dominated by gas drag we did not collect more data here.

\subsection{Twisting}

As noted above, this experiment is the only one where twisting on an nm-size contact can be studied, so far. Rotation around the vertical or around the direction of thermophoretic pull was observed on many aggregates before their break-up. That implies that a torque on the aggregates exists. There are again different cases to be distinguished, either the initial rotational acceleration is resolved from an aggregate at rest or a more or less constant rotation frequency is observed. In the first case, the accelerating torque can be measured directly. In the latter case, damping equals the acceleration and the damping can be deduced if the accelerating torque was measured before.

\begin{figure}
 \centering
 \includegraphics[width=.47\textwidth]{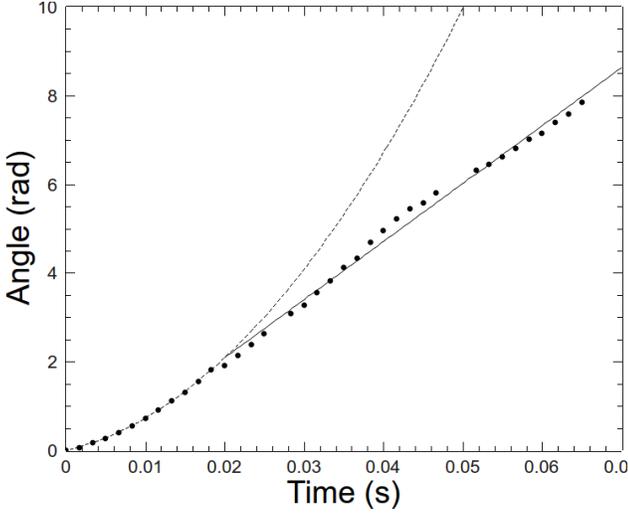}
 \caption{Angle of a selected aggregate pixel around the rotation axis.}
 \label{rotation}
\end{figure}

The value of $M_t$ is determined experimentally again using  equation \ref{torque} and \ref{inertia}. This time, the angular acceleration $\alpha$ corresponds to the aggregate rotation around a vertical axis though, and $r_i$ refers to the distance between the $i$-th aggregate pixel and the rotation axis. Since the images of the aggregates are 2D projections, the angular position of a pixel traced can be estimated by $\theta = \arcsin(x'/x)$, where $x$ is the maximum amplitude of the selected pixel on the horizontal axis and $x'$ the horizontal position at a given moment. The pixel's path can be plotted as a function of time as can be seen in Fig \ref{rotation}.

Considering that the position of the selected pixel describes a uniform accelerated rotation until a constant velocity is achieved, a parabolic fit gives the initial angular acceleration, e.g. $\alpha = 6300 \pm 200$ rad/s$^{-2}$ in the case of Fig. \ref{rotation}.

\begin{table*}[htbp]
\caption{Twisting torques, as measured, predicted by the theory at the break-up moment and at the moment when aggregates start to twist.}
\centering
\begin{tabular}{|c|c|c|c|c|c|}
\hline
Aggregate & \multicolumn{1}{l|}{$M_t$(exp) (Nm)} & \multicolumn{1}{l|}{$M_t$(theory) (N m)} & \multicolumn{1}{l|}{$M_t^{corr}$(theory) (N m)} & \multicolumn{1}{l|}{$\frac{M_t\textrm{(exp)}}{M_t \textrm{(theory)}}$} & \multicolumn{1}{l|}{$\frac{M_t\textrm{(exp)}}{M_t^{corr}\textrm{(theory)}}$} \\ \hline
2 & 3.97E-015 & 2.25E-17 & 2.80E-16 & 176.4 & 14.2\\ \hline
3 & 2.95E-015 & 6.69E-17 & 8.91E-16 & 44.1 & 3.31\\ \hline
4 & 9.94E-017 & 1.07E-18 & 1.14E-15 & 92.9 & 0.09\\ \hline
5 & 1.75E-016 & 8.61E-19 & 6.49E-16 & 203.2 & 0.27\\ \hline
8 & 1.47E-015 & 9.10E-18 & 4.45E-16 & 161.5 & 3.30\\ \hline
9 & 1.81E-016 & 1.53E-18 & 4.73E-16 & 118.3 & 0.38\\ \hline
10 & 4.20E-016 & 5.66E-19 & 1.13E-15 & 742.0 & 0.37\\ \hline
11 & 4.71E-017 & 1.69E-19 & 6.20E-16 & 278.0 & 0.08\\ \hline
12 & 3.39E-017 & 5.91E-20 & 9.17E-16 & 573.8 & 0.04\\ \hline
\end{tabular}
\label{twisting}
\end{table*}

\begin{figure}
 \centering
\includegraphics[width=.47\textwidth]{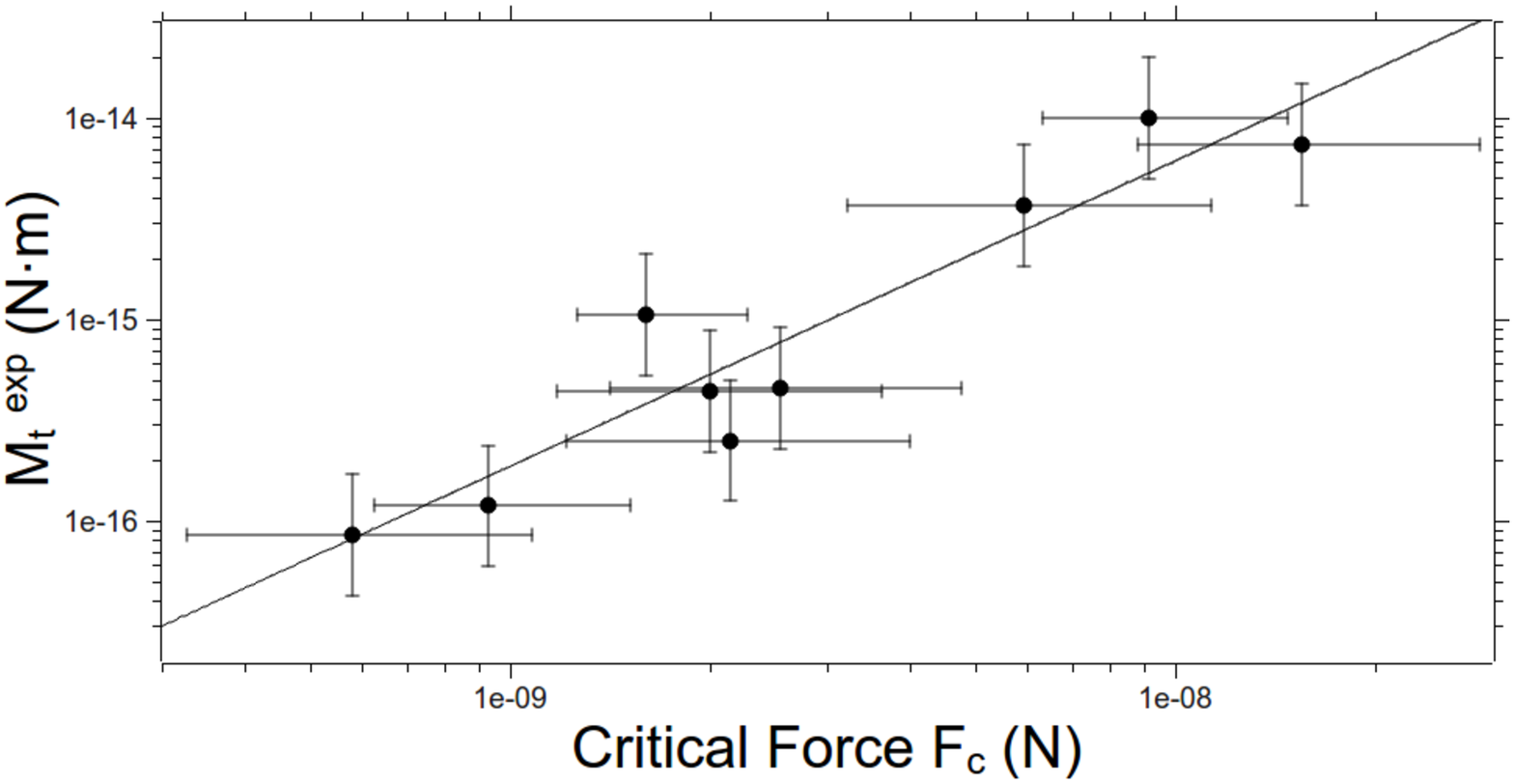}
 \caption{Twisting torque over break-up force. The solid line corresponds to linear dependence fitted to the data.}
 \label{mtvsfc}
\end{figure}

Fig. \ref{mtvsfc} shows the twisting torque depending on the break-up force. There seems to be a linear trend between twisting torque and break-up force. It has to be noted that twisting starts before break-up and the contact area will be larger at this time due to sublimation. We will consider this in the section below where we compare the results to the model. 

Measuring a finite torque once a particle starts a twisting motion implies that the friction is reduced suddenly. This is obviously similar to the macroscopic case where sliding friction is smaller than sticking friction. Otherwise rotation should start very slowly.

\subsection{Gas--grain coupling}

Coagulation via hit and stick events is the dominating growth process during the first steps of planetesimal formation. Relative velocities are generated by differences in gas--grain coupling or friction times. It is therefore important to know how different kinds of aggregates (e.g. fractal cluster--cluster aggregates) couple to the gas. If torques by radiation or gas drag act on an aggregate, it will rotate with a constant velocity eventually. This rotation might be related to alignment, and it might provide a significant part of the collision energy if two aggregates collide. The final rotation frequency is set by gas drag. Experiments are lacking for this. The observed interactions in our experiments are directly related to the interaction between gas and $\mu$m ice grains. A systematic study on cluster-cluster aggregates would provide basic information needed to detail the physics of particle motion and collisions in protoplanetary discs.

\begin{figure}
 \centering
\includegraphics[width=.47\textwidth]{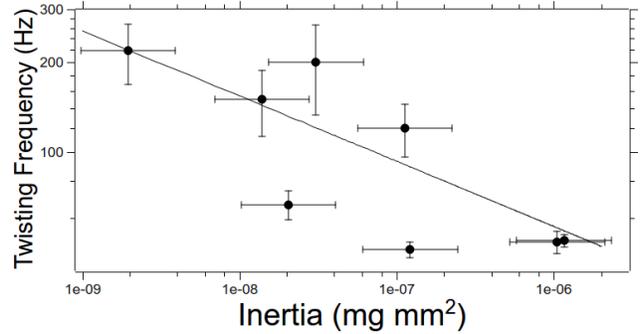}
 \caption{Twisting frequency measured for different aggregates.}
 \label{HzVSIn}
\end{figure}

Rotation and gas--grain coupling related to this is not easily measured for microscopic particles. \citet{vanEymeren2012} observed rotation frequencies of 1--100 Hz for 10--100 $\mu$m size ice aggregates freely levitated. This is comparable to the rotation frequencies measured here related to twisting. The setup therefore offers one way to study the interaction between gas and aggregates related to rotation. Especially, it allows us to determine coupling times. In Fig. \ref{HzVSIn} we show the rotation frequencies over aggregate moments of inertia. 

The equilibrium frequency is $\omega = \alpha \tau_{rot}$, and for typical angular accelerations of $10^4 \text{rad\:s}^{-2}$, we get a coupling time of $\tau_{rot} \sim$ 0.04 s at a pressure of $\sim$ 0.5 mbar. To set this in context, we compare this to the linear coupling time for spherical particles of the aggregate masses at the same pressure calculated by \citet{Blum1996a}:
\begin{equation}
 \tau_f=\varepsilon\frac{m}{\sigma_{\text{a}}}\frac{1}{\rho_{\text{g}} v_{\text{m}}}
 \label{}
\end{equation}
where $\varepsilon$ is a constant 0.68, $m$ the mass of the aggregate, $\sigma_{\text{a}}$ its cross-section, $\rho_{\text{g}}$ the gas density and $v_{\text{m}}$ the mean thermal velocity of the gas molecules. Introducing the parameters corresponding to our experimental conditions, the linear coupling time is of the order of $\tau_f \sim 0.10$ s. This is consistently of the same order and the quick estimate shows the ability of the setup to quantify these properties. Details are a separate topic and beyond this paper.

\section{Discussion}

\label{hundeleine}

How do the experimental results especially on twisting compare to existing models for macroscopic particles? The first thing to note is that a contact around which a continuous twisting rotation is induced is still holding if a pulling force is applied. Macroscopically, if a contact breaks to allow sliding or twisting such a particle would instantly lift off under a pulling force. This is not the case for the ice aggregates. This indicates that friction is indeed provided by stepwise motion of atoms from one potential well into the next one, and during twisting a contact only loses part of its sticking ability. This is the idea of the two extra terms for ice and iron in equation \ref{Mt} provided by \citet{Dominik1997}. The observation of a freely rotating contact might also occur if the particle was a fluid. \citet{johnston2012} show that water ice particles at 200 K behave like fluids if they are smaller than $\sim$1.4 nm. This is of the order of the size of some of the particles if we assume the highest value for $\gamma$ but is much smaller for the lowest $\gamma$. Also, rotation should not set in with a sudden jump in the case of a fluid connection unless the transition is exactly reached during sublimation. However, especially for the small particles, sublimation is significant and implies that twisting sets in at much larger sizes where particles would not be fluid. In any case, the largest particles studied are way larger and we assume that we can regard them as solid here.

With the given technique, we cannot resolve the contacting particles nor the contact area. Therefore, we use equation \ref{critforce} to deduce a reduced radius assuming a certain value of $\gamma$. This assumes that this equation still holds and should give a size estimate of the order of magnitude but otherwise is a formal parameter here. 

We then use equation \ref{a0} to estimate the equilibrium contact area $a_0$. The contact area under load $a$ may be substituted in all cases by $(a/a_0)a_0$. The ratio $a/a_0$ might vary between 0.63 and 1 (see Fig \ref{aa0}). However, we take $a/a_0$ = 0.63 in all cases here due to the proximity of the applied force to the critical force for break-up.

\subsection{Rolling}

An experimental torque for rolling motion was measured from the oscillations, but it proved to be of the same order as expected for a pendulum driven by thermophoresis of a displaced aggregate. This implies that the rolling torque due to the contact is much smaller. Assuming the validity of the theory on the small scale it can be estimated from equation \ref{Mr}. Values are plotted in table \ref{rolling} for a contact area at the time of break-up. They are a factor of $\sim 1500$ to $\sim 3\times 10^5$ smaller than the observed torque. If sublimation is considered, the contact area should be larger at earlier times and the torque due to the contact as well. This should decrease the amplitude of the oscillation at earlier times. However, no change is observed. Values for the rolling torque deduced from the damped case are consistent with gas drag, and while an order of magnitude smaller than the thermal gradient torques they are still two orders of magnitude larger than the modelled values.

All this indicates that the contact has an insignificant part in the measured rolling motions. The measurements therefore (only) provide an upper limit to the contact rolling torques. We estimate that we should be capable of seeing effects if the contact torques were on the 10 per cent level of the measured torque. This implies that the best guess for an upper limit for real torques as given in Table \ref{rolling} is a factor of 100 larger than the given model values, meaning that on the nanoscale the model is still consistent, but deviations cannot be excluded from the experiments.

\subsection{Twisting}

\begin{figure}
 \centering
 \includegraphics[width=.47\textwidth]{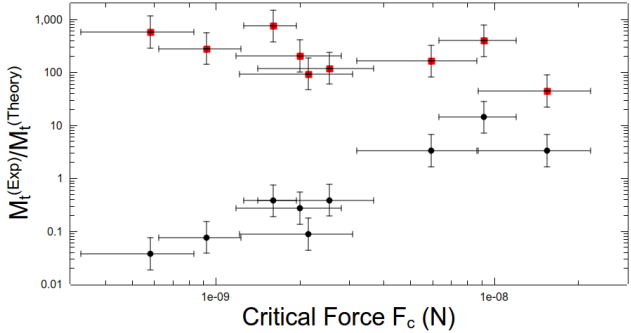}
 \caption{Comparison of experimental twisting torque and theoretical twisting torque. $\gamma=0.10 \text{J\:m}^2$. Red squares: uncorrected ratios; Black circles: twisting corrected ratios for sublimation to the time twisting sets in.}
 \label{Twistingratio}
\end{figure}

Using equation \ref{Mt} a theoretical value  for twisting can be deduced. We use $\gamma=0.1 \rm J\:m^2$ and the values provided by \citet{Dominik1997} $b=3.36$ \AA, $E=7\times 10^9 \text{N\:m}^2$, $G=2.8\times 10^9 \text{N\:m}^2$ and $\sigma=3.36$ \AA. The experimental and theoretical results of $M_t$ for different aggregates are shown in Table \ref{twisting} and the ratios are shown in Fig. \ref{Twistingratio}. The figures contain two data sets, one where the measured twisting torque is directly compared to the modelled one. The other (black circles) is corrected using equation \ref{Twistingratio}, considering that the sublimation changes the reduced radius in the time between onset of twisting and break-up.

Especially for the small break-up forces which go with small reduced radii, this has a large influence on the result. We note that this correction assumes that the reduced radius and contact areas change according to the simple sublimation model and that the dependence of the twisting torque on the contact area given by the model is correct. If so, this would suggest that the modelled torque overestimates smaller particles in contact and underestimates torques for the larger ones. A less extreme correction might reduce the ratio between measured torque and modelled torque somewhat less, but this is mere speculation so far. In that case, especially in view of the ratio at larger break-up forces which are less sensitive to corrections, the measured twisting torque would be about a factor of 10 larger than the modelled one.

Not considered yet is that the twisting torque is usually considered to increase linearly with the angle in the elastic regime and stays constant once the inelastic regime is reached. In that case twisting motion would set in smoothly. However, the twisting motion has a clear starting point and a finite initial acceleration. We note though that we only measure the decrease between maximum elastic torque and reduced torque while twisting. One might argue that we keep the terms related to the water which still lead to a bound system while twisting and the decrease in torque would be about half of the maximum torque at transition. This 'only' changes the ratio between experiment and model by a factor of 2 though, which is currently well within the uncertainty range.

We considered only a model with spherical contacts. This might not necessarily be the case in reality. Rather unshaped contact areas are a possibility as well \citep{Mo2010}. Twisting might change the shape and size of the area which would imply a possible jump in the twisting torque. This would fit to the observation that the rotation velocity can vary periodically (see Fig. \ref{rotation}). In some cases, the rotation almost comes to rest again in certain positions though this can also be explained by the influence of gravity on an asymmetric aggregate. However, an extreme of this idea of non-spherical contacts would be a scenario in which two separated contacts would exist some distance apart. These can provide a much larger torque. If one of these contacts then breaks or vanishes e.g. due to sublimation, the other contact would be too weak to oppose the thermal gradient torque and would start to be twisted  instantly with a finite acceleration. We cannot rule this out completely for all cases, but as the second contact also prevents rolling in a certain direction a particle should relax visibly by realignment upon break-up of this second contact towards the thermophoretic force. Such events occur but are not analysed here.

Another idea that cannot be ruled out a priori is that a contact initially is not just based on surface forces but rather sintered together. This might initially provide a larger torque and would explain a jump to a finite acceleration once the contact breaks. However, we observe aggregates to start twisting which readjusted around this contact before by rolling, aligning themselves to the thermophoretic pull. Therefore, such contacts would no longer be sintered.

If one of this possibilities were true, we would again measure an upper limit. This would not explain though why we see twisting aggregates which change their orbital velocity periodically in a steady way as seen in Fig. \ref{rotation}. This cannot be explained if a second contact vanished before or a sintered contact breaks, and this rather suggests that we really measure a twisting torque related to the contact.

We currently do not have any other theory at hand to compare our experimental data to. We also cannot resolve the size of the contact area directly. We therefore have to imply the size from an equation without knowing if it applies.  However, if we use this model, it still provides an analytic expression how to calculate the forces and torques in kind of a self-consistent way if we consider a potential correction factor deduced from experiments. Certainly, a theory e.g. molecular dynamics describing the contact forces at this small scale including torques would be desirable but is not yet available even though current work is pointing in this direction  \citep{Tanaka2012}.

\section{Summary and conclusions}

\label{wassersack}

The motivation behind this work is the need to understand contacts between grains in astrophysical environments. Much work on this has already been carried out as outlined in the introduction. In collisions with energy low enough, individual grains in an aggregation process stay intact. The contacts between them especially if the forces are only surface forces (surface tension, van der Waals forces and dipole interactions) are the weak connections. There are four processes related to changes in the contact: complete break-up, sliding, rolling and twisting. The importance of each process varies for different overall aggregate structures and porosities. Measurements for large ($\mu$m-size and larger) silicate (dust) particles exist with respect to break-up and rolling, and the equations given above are valid and can be applied. 

What is lacking is the knowledge on sliding and twisting and an extrapolation to the nm-scale as e.g. interstellar grains are often supposed to be built from nm-monomers \citep{Mathis1986,Dwek2004}. Also lacking is the knowledge on the contact physics of ice. While sliding, rolling and break-up can well be studied by AFM on silicates, this technique has not yet provided data on twisting on the small size scale. What we accomplished in general by our new setup can be listed as follows.

\begin{itemize}
\item First of all we developed a new method to determine small forces between
nm-size particles and especially one which allows twisting to be observed.

\item This technique can be used to study thermophoretic forces on aggregates.

\item This method can also be used to study rotational gas--grain coupling.

\item Most of all the method can relate break-up forces to rolling and twisting torques.

\item The experiments so far provide upper limits on the torques which nm-ice contacts
can provide to oppose rolling. These upper limits are a factor of 100--1000 larger than the existing contact model given by \citet{Dominik1997} scaled down to nm-size. As these are upper limits, the torques are still consistent with the model, but we cannot exclude the torques to be a factor 100 to 1000 larger.

\item We find torques opposing twisting which are up to a factor of 10 larger than the model would predict. In our view of the data these are not upper limits and a correction factor of 10 might be appropriate for nm-ice-particles if the model -- in lack of any other model -- is to be kept.

\item Qualitatively, it might also be worth noting that ice particles can
rotate around their contact 'freely' even if the contacting particles are pulled apart.

\end{itemize}

\subsection{Astrophysical application}
Twisting, sliding, rolling and breaking are of fundamental importance to understand aggregation in astrophysical settings. This ranges from interstellar aggregates to aggregates in protoplanetary discs. The fact that ice aggregates are more robust against twisting implies that they are less likely to be restructured. If the evolution of large particles currently discussed for coreshine e.g. by \citet{Steinacker2010} and \citet{Pagani2010} is modelled this might be a fact to consider. \citet{Seizinger2012} modelled aggregate compression and in order to explain experimental results on compression they had to stiffen the contacts, i.e. adding a factor to the sliding force.

As sliding and twisting are based on the same mechanisms the higher values measured in our experiments might support this ad hoc modification by \citet{Seizinger2012}. In general, ice and dust aggregates of small grains growing their way to planetesimals would have a higher porosity at a given size. With such changes, transitions in planetesimal formation would shift or change. From our very basic results, we can only speculate so far, but the transitions between sticking and bouncing or fragmentation might shift to larger sizes if smaller particles and/or ice are considered \citep[Kelling et al., submitted]{Teiser2009, Windmark2012a, Kelling2013}. This might allow particles to reach a size where they get susceptible to concentration in turbulent discs and subsequent gravitational collapse \citep{Chiang2010}. In this sense, the results can shift the picture of planetesimal formation quite a bit.

Thermal creep provides a means to produce a gas flow around and through a particle. If sublimation takes place in a dry environment, a large aggregate might be eroded from the outside as well as the inside by the gas flow. Therefore, a similar setup might also be used to observe how large highly porous ice aggregates might evolve in a gas flow, but this is a topic apart from contact physics.

\subsection{Future experiments}

So far we used ice particles but similar experiments should allow us to quantify contact forces between metallic nm-particles. However, as the ice experiments make use of sublimation in the present version, the thermophoretic force would need to be varied. This can be done by changing ambient pressure and temperature gradient. The effect of such variations have not been exploited systematically yet and are also a possible road to decrease the gas damping in oscillations. As temperature is important, a faster control is most desirable e.g. to prevent a rotating dust aggregate from further sublimation and break-up for long-term studies. Also sublimation itself for different parameter ranges has to be studied in more detail. This is currently one of the largest uncertainties for correcting between times of break-up and times of twisting

One idea to better discern contact dynamics from external forcing is resonant excitation, which is currently being tested. Oscillating the substrate with variable frequency should allow us to determine the rolling force of a contact even in view of larger thermophoretic torques. In general, analysis of the mass and motion of the ice aggregates can be improved by generating ice aggregates which are better defined (e.g. fractal aggregates) and observing them with better time and spatial (and 3D) resolution. A still somewhat speculative perspective is the construction of well-defined microprobes which can be attached to the ice particles under consideration and be subjected to external electrical or magnetic forces and torques.

If force balances can be adjusted accurate enough, this might also allow the study of dust particles in the future.

However, in the near future we aim at quantifying the contact dynamics of water ice particles to higher accuracy including twisting and break-up but also rolling which is currently only known in large bounds.

\section{Acknowledgements}
G A is funded within the European Commission's Seventh Framework Programme (FP7/ 2007-2013 under grant agreement No. 238258). We thank the anonymous referee for the review of our manuscript.

\bibliographystyle{mn2e}
\bibliography{biblio}

\label{lastpage}

\end{document}